\documentclass[11pt]{article}

\usepackage{fullpage}
\usepackage{hdb_macros}
\usepackage[draft,multiuser,inline,nomargin]{fixme}
\fxusetheme{color}

\title{Range Avoidance for Constant-Depth Circuits:\\ Hardness and Algorithms}
 \author{
    Karthik Gajulapalli
     \thanks{Georgetown University. Email: \texttt{kg816@georgetown.edu}.}
     \and
 	Alexander Golovnev
 	\thanks{Georgetown University. Email: \texttt{alexgolovnev@gmail.com}.}
 	\and
 	Satyajeet Nagargoje
 	\thanks{Georgetown University. Email: \texttt{satyajeetn2012@gmail.com}.}
 	\and
 	Sidhant Saraogi
 	\thanks{Georgetown University. Email: \texttt{ss4456@georgetown.edu}.}
 }
\date{}

\begin{document}
\maketitle
\begin{abstract}

Range Avoidance (\Avoid{}) is a total search problem where, given a Boolean circuit $\Ck\colon\{0,1\}^n\to\{0,1\}^m$, $m>n$, the task is to find a $y\in\{0,1\}^m$ outside the range of~$\Ck$. For~an integer $k\geq 2$, $\NC^0_k$-\Avoid{} is a special case of \Avoid{} where each output bit of~$\Ck$ depends on at most~$k$ input bits. While there is a very natural randomized algorithm for \Avoid{}, a deterministic algorithm for the problem would have many interesting consequences. Ren, Santhanam, and Wang (FOCS 2022) and Guruswami, Lyu, and Wang (RANDOM 2022) proved that explicit constructions of functions of high formula complexity, rigid matrices, and optimal linear codes,  reduce to $\NC^0_4$-\Avoid{}, thus establishing conditional hardness of the $\NC^0_4$-\Avoid{} problem. 
On the other hand, $\NC^0_2$-\Avoid{} admits polynomial-time algorithms, leaving the question about the complexity of $\NC^0_3$-\Avoid{} open.

We give the first reduction of an explicit construction question to \mbox{$\NC^0_3$-\Avoid{}}. Specifically, we prove that a polynomial-time algorithm (with an $\NP$ oracle) for $\NC^0_3$-\Avoid{} for the case of $m=n+n^{2/3}$ would imply an explicit construction of a rigid matrix, and, thus, a super-linear lower bound on the size of log-depth circuits.

We also give deterministic polynomial-time algorithms for all $\NC^0_k$-\Avoid{} problems for $m\geq n^{k-1}/\log(n)$. Prior work required an $\NP$ oracle, and required larger stretch, $m \geq n^{k-1}$.

\end{abstract}
\pagenumbering{roman}
\thispagestyle{empty}
\newpage
\setcounter{tocdepth}{2}
\newpage
\pagenumbering{arabic}

\section{Introduction}

The Range Avoidance (\Avoid{}) problem is: given a Boolean circuit $\Ck\colon \{0, 1\}^n \to \{0, 1\}^{m}$ for some \emph{stretch} $m>n$, find an element $y\in\{0,1\}^m$ outside the range of $\Ck$. By the pigeonhole principle, such a~$y$ always exists. This problem was first introduced by Kleinberg, Korten, Mitropolsky, and Papadimitriou~\cite{kleinberg2021total} as a complete problem for the class $\APEPP$ (Abundant Polynomial Empty Pigeonhole Principle). Informally, $\APEPP$ contains total search problems where the existence of a solution follows via the union bound (such as Shannon's classical proof that most functions require circuits of exponential size). 

Korten~\cite{korten2022hardest} proved that a deterministic algorithm for $\Avoid{}$ would imply explicit constructions of objects that are central to the field of computational complexity, and would resolve several long-standing open problems. Such objects include functions of high circuit complexity, rigid matrices, pseudorandom generators, and Ramsey graphs. The key idea is that there is a succinct way to encode all ``easy'' objects (such as descriptions of functions of low circuit complexity) in the input space of a small circuit that acts as a decoder.  Then a solution to the \Avoid{} problem yields a ``hard'' object (such as a function of high circuit complexity), implying an explicit construction. 
In fact, the aforementioned works~\cite{kleinberg2021total,korten2022hardest} showed that even a deterministic algorithm \emph{with an $\NP$ oracle} solving \Avoid{} in polynomial time would lead to breakthrough results in complexity theory.

The only known deterministic algorithm for $\Avoid{}$ is the trivial brute force algorithm running in time $2^n \cdot \poly(n, |\Ck|)$.\footnote{This trivial algorithm is (conditionally) tight for a related problem studied in~\cite{kleinberg2021total}, where the range of $\Ck$ has size much smaller than $2^{n+1}$, and is given by a circuit computing a function from $[N]$ to~$[M]$. \cite{kleinberg2021total} gives a deterministic reduction from $\SAT$ on~$n$ variables to $\Avoid{}$ for a circuit $\Ck\colon[2^n]\to[2^n+2^{o(n)}]$ running in subexponential time. Thus, under the Exponential Time Hypothesis~\cite{IPZ98,IP99}, this problem does not admit deterministic (and randomized) algorithms running in time $2^{o(n)}$.} No better algorithms are known for \Avoid{} even if the algorithm is allowed to use an $\NP$ oracle. On the other hand, using both an $\NP$ oracle and randomness, one can solve $\Avoid{}$ in polynomial time: Pick a random string $y\in\{0,1\}^m$, and simply check if $y\in\Range(\Ck)$ using the $\NP$ oracle. This shows that $\Avoid\in\FZPP^\NP$, and, using the standard trick of simulating randomness by non-uniformity, $\Avoid\in\FP^\NP/\cc{Poly}$.\footnote{Here, the complexity classes $\FP,\FE,\FZPP$ are simply the functional analogs of the decision classes $\P, \E, \ZPP$.} Interestingly, for the class of polynomial-time algorithms with an $\NP$ oracle, the $\Avoid{}$ problem is equally hard for all values of stretch $n+1\leq m \leq \poly(n)$~\cite{kleinberg2021total}.

For a class of circuits~$\mathcal{C}$, the $\mathcal{C}$-\Avoid{} problem is a special case of \Avoid{} where each output is computed by a circuit from~$\mathcal{C}$. Recent works by Ren, Santhanam, and Wang~\cite{ren2022range} and Guruswami, Lyu, and Wang~\cite{guruswami2022range} proved that efficient algorithms for $\mathcal{C}$-\Avoid{} even for certain simple circuit classes $\mathcal{C}$ would be sufficient for getting various explicit constructions.  Later, Chen, Huang, Li, and Ren~\cite{CHLR23} re-derived the best known lower bounds against $\ACC^0$ circuits from an efficient algorithm for a certain $\mathcal{C}$-\Avoid{} problem. While this suggests that designing efficient algorithms for \Avoid{} problems is a promising approach to various explicit construction questions, the work of Ilango, Li, and Williams~\cite{ILW23} proves barriers for designing polynomial time algorithms under certain cryptographic assumptions.

Let $\NC^1$ denote the class of Boolean fan-in-$2$ circuits of depth $O(\log(n))$, and $\NC^0_k$ denote the class of Boolean functions where each output depends on at most~$k$ inputs for a constant~$k$. \cite{ren2022range}~used perfect encodings of~\cite{IK00,IK02,applebaum2006cryptography} to reduce $\NC^1$-\Avoid{} to $\NC^0_4$-\Avoid{} in polynomial time. Consequently, \cite{guruswami2022range} reduced most of the aforementioned explicit constructions in~\cite{korten2022hardest} (and several new ones!)  to $\NC^1$-\Avoid{}, and, thus, to $\NC^0_4$-\Avoid{}. In particular, polynomial-time \emph{deterministic} algorithms (even with an $\NP$ oracle) for $\NC^0_4$-\Avoid{} would now imply breakthrough results in complexity theory. 

\cite{guruswami2022range} gave a polynomial-time algorithm solving $\NC^0_2$-\Avoid{} for any stretch $m\geq n+1$. As mentioned above, $\NC^0_4$-\Avoid{} might be hard to solve efficiently. This leaves the question about the complexity of $\NC^0_3$-\Avoid{} open. 

\begin{openproblem}[\cite{guruswami2022range}]\label{op:hardness}
Can we reduce explicit construction problems to solving \mbox{$\NC^0_3$-\Avoid{}}? Or can we solve $\NC^0_3$-\Avoid{} in polynomial time?
\end{openproblem}

Unlike the case of the general \Avoid{} problem, $\NC^0_k$-\Avoid{} may be much easier for large values of the stretch~$m$. Indeed, on one hand, $\NC^0_k$-\Avoid{} for small stretch $m=n+o(n)$ is capable of encoding hard explicit construction problems~\cite{ren2022range,guruswami2022range}. On the other hand, $\NC^0_k$-\Avoid{} for $m=\Omega(n^k)$ is easily solvable in polynomial time: since the number of distinct functions depending on at most $k$ out of $n$ inputs is $O(n^k)$, every such instance of the problem must have two outputs computing identical functions.  Assigning different values to these outputs solves $\NC^0_k$-\Avoid{}.

\cite{guruswami2022range} presented an algorithm solving $\NC^0_k$-\Avoid{} for stretch $m\geq\Omega(n^{k-1})$ in polynomial time with an $\NP$ oracle.\footnote{The algorithm of~\cite{guruswami2022range} does not use the full power of $\FP^\NP$: it outputs a hitting set $H\subseteq\{0,1\}^m$ such that for every $\NC^0_k$ function~$\Ck$, at least one point $y\in H$ is outside the range of $\Ck$. Only then the algorithm looks at the input function and finds a solution $y\in H$ using the $\NP$ oracle.} This improvement on the trivial algorithm suggests a natural question of whether one can solve $\NC^0_k$-\Avoid{} for even smaller values of stretch~$m$. 

\begin{openproblem}\label{op:alg}
Design a polynomial-time algorithm (with an~$\NP$ oracle) solving $\NC^0_k$-\Avoid{} with $n$ inputs and stretch $m=o(n^{k-1})$ for $k\geq3$.
\end{openproblem}

\subsection{Our Results}
The classical result of Shannon~\cite{Shannon49} shows that most Boolean functions of $n$ variables require Boolean circuits of exponential size. Despite that, the best known lower bound on the size of a circuit (or even a circuit of logarithmic depth, i.e., $\NC^1$) for a function in~$\P$ (or even $\E^\NP$) is $3.1n-o(n)$ proven by Li and Yang~\cite{LY22}. A central problem in circuit complexity is to prove a super-linear lower bound on the number of gates of $\NC^1$ circuits computing an explicit function~\cite[Frontier~3]{valiant1977graph, AB2009}.

Similarly, for the class of \emph{linear} $\NC^1$ circuits---$\NC^1$ circuits where each gate computes the XOR (or its negation) of its two inputs---no super-linear lower bound on the complexity of an explicit linear map $M\in\F_2^{n\times n}$ is known. The best lower bound against linear circuits is $3n-o(n)$ proven by Chashkin~\cite{chashkin}.

In our first result (\cref{thm:rrlb}), we answer~\cref{op:hardness} by showing that a polynomial-time algorithm for $\NC^0_3$-\Avoid{} would imply an explicit construction of a map requiring linear $\NC^1$ circuits of super-linear size (thus, demonstrating the hardness of $\NC^0_3$-\Avoid{}).

\begin{thm}\label{thm:thm1}
 An $\FP$ (resp. $\FP^\NP$) algorithm for $\NC^0_3$-\Avoid{} with stretch $m=n+O(n^{2/3})$ implies an explicit construction of a linear map in $\FP$ (resp. $\FP^\NP$) that cannot be computed by linear $\NC^1$ circuits of size $o(n\log\log(n))$.
\end{thm}

Our proof of \cref{thm:thm1} first reduces an explicit construction of a rigid matrix to  $\NC^0_3$-\Avoid{} (\cref{thm:rrlb}). A matrix $M\in\F_2^{n\times n}$ is called $(r,s)$-rigid if it cannot be written as a sum $M=L+S$ of a rank-$r$ matrix $L$ and a matrix $S$ with at most $s$ non-zeros per row. In a seminal work, Valiant~\cite{valiant1977graph} introduced an approach for proving super-linear lower bounds on the size of linear $\NC^1$ circuits via matrix rigidity. Valiant proved that an $(\eps n, n^\eps)$-rigid matrix $M\in\F_2^{n\times n}$ for any constant $\eps>0$ requires linear $\NC^1$ circuits of size $\Omega(n\log\log(n))$. \cref{thm:thm1} now follows straightforwardly as a corollary of \cref{thm:rrlb}.

The best known constructions of rigid matrices do not yet achieve the parameters sufficient for Valiant's circuit lower bound. \cite{f93,pr94,sss97} construct an $\left(r, \Omega \left (\frac{n}{r}\log(\frac{n}{r}) \right)\right)$-rigid matrix in polynomial time, \cite{goldreich2016matrix} gives an $\left(r, \Omega \left(\frac{n^2}{r^2\log(n)}\right)\right)$-rigid matrix in time $2^{O(n)}$ for $r\geq\sqrt{n}$, and \cite{ac19,bhpt20} give $(2^{\eps\log(n)/\log\log(n)},\Omega(n))$-rigid matrices in polynomial time with an $\NP$ oracle. However, even an $\FP^{\NP}$ algorithm for $\NC_3^0$-\Avoid{} with stretch $m = n + n^{12/17 - \eps}$ for any constant $\eps > 0$ would already improve on these known constructions of rigid matrices. 

In fact, we reduce the problem of constructing explicit rigid matrices to a problem that we call degree-$2$-\Avoid{}, where each output computes a degree-$2$ polynomial of the inputs. Following Ren, Santhanam, and Wang's approach~\cite{ren2022range}, this problem can be reduced to $\NC^0_3$-\Avoid{} using the perfect encoding scheme of Applebaum, Ishai, and Kushilevitz~\cite{applebaum2006cryptography}.\footnote{Ren, Santhanam, and Wang use the following definition of perfect encodings: A function $\widehat{f}$ is a perfect encoding of a function $f$ if there exists a polynomial time algorithm $\mathrm{Dec}$ such that for all $x, y$: $\mathrm{Dec}(y) = f(x) \iff \exists r, y = \widehat{f}(x, r)$.} 

On the algorithmic side, we make partial progress towards resolving \cref{op:alg}. We first give a simple deterministic polynomial-time algorithm for $\NC^0_3$-\Avoid{} for stretch $m\geq\binom{n}{2}/3+2n$ (presented in \cref{sec:appendix}). This algorithm already improves on the best known algorithm for $\NC^0_3$, as it does not use an $\NP$ oracle. Then, in \cref{thm:good-stretch} we extend this algorithm to solve $\NC^0_k$-\Avoid{}  for all constant~$k$. Recall that the current best algorithms for this problem solve the case where $m\geq\Omega(n^{k-1})$ in polynomial time using an $\NP$ oracle~\cite{guruswami2022range}. We improve this result in two directions: our algorithm does not use an $\NP$ oracle, and it works in polynomial time for stretch $m\geq n^{k-1}/\log(n)$. 

\begin{thm}\label{thm:thm2}
 There is a deterministic polynomial-time algorithm that solves the $\NC^0_k$-\Avoid{} problem with $n$ inputs and stretch $m$ for every $k\geq3$ and $m\geq n^{k-1}/\log(n)$.
\end{thm}

\subsection{Proof Overview}
\paragraph{Hardness of $\NC^0_3$-\Avoid{}.}

Valiant~\cite{valiant1977graph} proved that linear $\NC^1$ circuits with a linear number of gates can only compute non-rigid linear maps $M\in\F_2^{n\times n}$, i.e., maps $M$ that can be written as a sum $M = Q + S$, where $\rank(Q)\leq \epsilon n$ and each row of $S$ has at most $n^\delta$ ones in it. For the rest of the section, our non-rigid matrices can be written as the sum of a matrix with rank $\leq n/10$ and a matrix with row sparsity at most $n^{0.1}$. Therefore, constructing a rigid matrix would imply a super-linear lower bound on the size of linear $\NC^1$ circuits computing~it.

To reduce an explicit construction of an $n \times n$ rigid matrix to solving an instance of $\NC^0_3$-$\Avoid$, we design an $\NC^0_3$ function $f\colon \{0,1\}^{p(n)} \to \{0,1\}^{n^2}$, for some polynomial $p(n) < n^2$, such that for every non-rigid matrix $M\in\{0,1\}^{n\times n}$, there exists $x \in \{0, 1\}^{p(n)}$ satisfying $f(x) = M$. Now, any solution $M' \in\{0,1\}^{n \times n}$ to the $\NC^0_3$-\Avoid{} problem for the function $f$ must be a rigid matrix.

Before constructing such an $\NC^0_3$ function~$f$, we first design a function $g\colon\F_2^{n^2/2}\to\F_2^{n^2}$, where each output bit of~$g$ is a degree-$2$ polynomial of the inputs, and the range of $g$ contains all non-rigid matrices. A solution to the degree-$2$-\Avoid{} problem for the function~$g$ would give us a rigid matrix. Following \cite{ren2022range}, we can then apply a perfect encoding scheme~\cite{IK00,IK02,applebaum2006cryptography} to $g$, and obtain an $\NC^0_3$ function~$f$, as required (see \cref{lem:degree-encoding}). Effectively, this reduces solving \Avoid{} on~$g$ to solving \Avoid{} on~$f$. 

Now we construct a degree-$2$ function~$g\colon\F_2^{n^2/2}\to\F_2^{n^2}$ whose inputs encode all non-rigid matrices, i.e., for every non-rigid matrix $M$, there is an $x\in\{0,1\}^{n^2/2}$ such that $f(x)=M$. A non-rigid matrix~$M$ can be written as $M=LR+S$, where $L,R^T \in \mathbb{F}_2^{n \times n/10}$, and each row of $S$ contains at most $n^{0.1}$ ones. The first $n^2/5$ inputs of the function~$g$ will correspond to the elements of $L$ and~$R$. Note that every entry of $LR$ is a degree-$2$ function of the entries of $L$ and $R$ since it just computes the inner product of a row in~$L$ and a column in~$R$. Now, for each $n^{0.1}$-sparse row of $S$, we show how to encode it using $n^{0.6}$ inputs and a degree $2$ function. Repeating this procedure for each row of~$S$ will finish the proof.

We interpret the $n^{0.1}$-sparse row with $n$ entries as a $\sqrt{n} \times \sqrt{n}$ matrix $A$. Since $A$ has at most $n^{0.1}$ non-zero entries, $\mathrm{rank}(A) \leq n^{0.1}$. It can be written as a product $A = BC$ where $B, C^T \in \mathbb{F}_2^{\sqrt{n} \times n^{0.1}}$. Therefore, there is a degree-$2$ function $h$ that takes as input $B, C$ of size $2n^{0.6}$ and outputs the sparse matrix $A$.

The presented encoding of $s$-sparse vectors in $\F_2^n$ is only non-trivial for $s<\sqrt{n}$ (as otherwise the number of inputs of~$h$ exceeds the number of outputs). As a result, we cannot encode the entire matrix $S$ as an $n^{1.1}$-sparse vector in $\F_2^{n^2}$. However, for Valiant's approach of proving circuit lower bounds, we can assume that $S$ is $n^{0.1}$-\emph{row} sparse.\footnote{Alternatively, one can argue by Markov's inequality that if a matrix $M$ cannot be written as a sum of rank-$r$ and $n^{0.1}$-\emph{row} sparse matrices, then $M$ also cannot be written as a sum of rank-$2r$ and $rn^{0.1}$-\emph{globally} sparse matrices.} Thus, we can separately encode each row of $S$ using only $O(n^{0.6})$ inputs to obtain an encoding of $S$ in $O(n^{1.6})$ bits. In \cref{lem: rigid_sparse_lemma}, we will demonstrate how to accommodate slightly better sparsity parameter when we are allowed higher (but still constant) degree $d$ for the encoding function. 

\paragraph{Simple algorithm for $\NC^0_3$-\Avoid{}.}
We start with a short description of a simple deterministic polynomial-time algorithm for $\NC^0_3$-\Avoid{} for stretch $m\geq\binom{n}{2}/3+2n$ (presented in \cref{sec:appendix}). This algorithm already improves on the best known algorithm for $\NC^0_3$-\Avoid{}, as our algorithm does not use an $\NP$ oracle.

If we had a $\#\SAT$ oracle, then we could solve  $\NC^0_3$-\Avoid{} even for stretch $m=n+1$. Our algorithm would iteratively find constant assignments to each of the first $n$ outputs to minimize the number of inputs that map to the current (partial) output assignments. Throughout our exposition, we say such inputs are \textit{consistent with the (partial) output assignments}. Before we describe our algorithm, it is important to note that we always fix circuit outputs to constant assignments. At each iteration, we would use the $\#\SAT$ oracle to find the output assignment that reduces the size of the input set by at least half. After fixing the first~$n$ outputs, we still have at least $m-n\geq1$ unassigned outputs, and only one input point $x\in\F_2^n$ that is consistent with the previously assigned output bits. This allows us to find an assignment of the $(n+1)$-th output bit such that the string specified by the output bits lies outside the range of the circuit.

Unfortunately, solving $\#\SAT$ (even approximately) is hard for this class of multi-output circuits. In the absence of an efficient $\#\SAT$ algorithm, our algorithm maintains an affine subspace $\AS$ that contains all inputs from $\F_2^n$ that are consistent with the current partial assignment ($\AS$ may also contain inputs that are not consistent with the current partial assignment). We carefully set output values so that at each iteration, we reduce the dimension of $\AS$ by at least one. This way, after $n+1$ steps we will find a solution to the $\NC^0_3$-\Avoid{} problem. However, our algorithm can only work when the stretch is $m \geq \Omega(n^2)$. 

Without loss of generality, we assume that each output reads exactly three input bits. At each iteration the number of currently unassigned outputs is $>\binom{n}{2}/3$. This allows us to find a pair of outputs $y_1$ and $y_2$ that share a pair of input variables.\footnote{Each output sees $3$ pairs of input bits, giving a total of $3m>\binom{n}{2}$ pairs. By the pigeonhole principle, at least one pair of inputs appears in two outputs.} Say, $y_1=f_1(x_1,x_2,x_3)$ and $y_2=f_2(x_2,x_3,x_4)$. We will find a constant assignment to $y_1$ and $y_2$ that reduces the dimension of the affine subspace $\AS$.

Note that there are 16 assignments to $(x_1,x_2,x_3,x_4)$ and four different values of $(y_1, y_2)$. Therefore, there is a way to assign $y_1=c_1, y_2=c_2$ such that at most four points $(x_1,x_2,x_3,x_4)$ map to these values of the outputs. Note that there always exists a hyperplane $H$ containing any four points in $\F_2^4$. Let $\AH$ be the affine subspace obtained by extending $H$ to all $n$ inputs. Then, $\AS\cap \AH$ gives us an affine subspace containing all inputs consistent with the assignment $y_1=c_1, y_2=c_2$.

If $\AS \nsubseteq \AH$, then we have an assignment of two outputs that reduces the dimension of our affine subspace as $\dim(\AS\cap \AH)<\dim(\AS)$. Otherwise, if $\AS \subseteq \AH$, then instead of considering  all 16 assignments to the inputs $(x_1,x_2,x_3,x_4)$, we can restrict our attention to at most~8 such assignments that belong to $H$, which makes the problem only easier (as we show in \cref{thm:apx}).

\paragraph{Better algorithm for $\NC^0_k$-\Avoid{}.}
The main bottleneck of this simple algorithm is that it maintains an affine subspace that must contain all consistent inputs. While affine subspaces are easy to work with, they are not expressive enough to accurately describe all inputs that are consistent with an arbitrary partial assignment. In order to improve the previous algorithm, we will maintain a more expressive structure than an affine subspace---a union of affine subspaces. Below we sketch our approach to solving $\NC^0_3$-\Avoid{} with stretch $m\geq \Omega(n^2/\log(n))$. \cref{thm:good-stretch} generalizes this to solving $\NC^0_k$-\Avoid{} problems with stretch $m \geq \Omega(n^{k-1}/\log(n))$ for all values of~$k$. 

Again, without loss of generality, we assume that each output depends on exactly three inputs. Consider a bipartite graph, where the left vertices correspond to~$n$ inputs, the right vertices correspond to~$m$ outputs, and an input-output pair $(x_i,y_j)$ is connected by an edge if the output~$y_j$ depends on the input $x_i$. First we select $t= 3n^2/m$ highest-degree inputs $I=\{x_1,\ldots,x_t\}$. Their neighborhood must contain at least $3n$ distinct outputs $O=\{y_1,\ldots,y_{3n}\}$.\footnote{Since the number of outputs is $m$, and each output has degree~$3$, the number of edges in the graph is~$3m$, and the average degree of an input is $3m/n$. The $t$ highest-degree inputs then have total degree at least $3mt/n$, and must be connected to at least $mt/n=3n$ distinct outputs.} Let $\Ck$ be the sub-circuit defined on outputs from~$O$ and their corresponding inputs. Now, we will find a $y\in\F_2^{3n}$ outside $\Range(\Ck)$.

First, consider all $2^t$ assignments to the inputs in $I=\{x_1,\dots,x_t\}$, resulting in circuits $\Ck_1,\ldots,\Ck_{2^t}$. Since every output in $O$ is connected to at least one input from $I$, fixing an assignment to the inputs $I$ reduces each $\Ck_i$ to an $\NC^0_2$ circuit. In a way, we have reduced $\NC^0_3$-\Avoid{} to an OR of $2^t$ instances of $\NC^0_2$-\Avoid{}: we need to find a $y\in\{0,1\}^{3n}$ outside the ranges of all the $\Ck_i$'s. For each circuit $\Ck_i$, we will maintain an affine subspace $\AS_i$ containing all inputs consistent with the current partial assignment of the outputs. 

Our algorithm works by iteratively fixing the output bits from~$\{y_1,\ldots,y_{3n}\}$ such that at each step the total number of points in the (disjoint) union of the affine subspaces $\AS_i$ is reduced by a constant factor, eventually making all the subspaces empty. We observe (in \cref{lem:nc02-avoid}) that for any affine subspace $\AS_i$, one of the assignments $y_i=0$ or $y_i=1$ always reduces the dimension of~$\AS_i$ by one. Therefore, by picking the ``best'' assignment $y_i=c$ across all the subspaces $\AS_i$, we can reduce the size of the union of such affine subspaces by a constant factor of $4/3$. Repeating this procedure for $\log_{4/3}(2^n)+1<3n$ steps finishes the proof.

\subsection{Open Problems}
Our work motivates several natural questions about the complexity of $\NC^0_3$-\Avoid{} and \mbox{degree-$2$-\Avoid{}}. We reduce explicit constructions of rigid matrices to solving \mbox{degree-$2$-\Avoid{}}, and then $\NC^0_3$-\Avoid{},  with appropriate stretch. 
\begin{openproblem}
    Can other explicit construction questions be reduced to $\NC^0_3$-\Avoid{} or \mbox{degree-$2$-\Avoid{}}?
\end{openproblem}
Particularly, we suspect that the construction of linear and list-decodable codes with optimal parameters~\cite{guruswami2022range} might be good candidates for these reductions.

Using the encoding of ~\cite{applebaum2006cryptography} in the reduction from degree-$2$-\Avoid{} to $\NC^0_3$-\Avoid{} almost always decreases the required stretch to $m = n + o(n)$ (as highlighted in \cref{sec:hardness}). It would also be interesting to find a more efficient encoding or reduction from degree-$2$-\Avoid{} to $\NC^0_3$-\Avoid{}. This could potentially increase the stretch for $\NC^0_3$-\Avoid{} required to obtain explicit constructions thereby making the problem easier. 

 \begin{openproblem}
     Can we construct a more efficient reduction from degree-$2$-\Avoid{} to $\NC_0^3$-\Avoid{}?
 \end{openproblem}

We believe degree-$2$-\Avoid{} might be of independent interest since it allows for a larger stretch. For example, for improved constructions of rigid matrices, it suffices to solve degree-$2$-\Avoid{} for super-linear stretch $m\geq n^{12/11-\eps}$ for a constant $\eps>0$. 
In fact, \mbox{degree-$2$-\Avoid{}} is easy to solve when the stretch is $m \geq n^2$. Note that there are at most $\binom{n}{2}$ unique degree-$2$ monomials on $n$ variables. If $m \geq n^2$, then we can replace each unique monomial with a new variable. As a result, we will have $m$ linear functions in $< m$ variables. We can solve $\Avoid{}$ on this linear function instance by a dimension reduction strategy similar to the one outlined in the previous section.

\begin{openproblem}
    Are there algorithmic techniques to solve \mbox{degree-$2$-\Avoid{}} that do not use a reduction to  
 \mbox{$\NC^0_3$-\Avoid{}}? 
\end{openproblem}

For the $\NC^0_3$-\Avoid{} problem, our algorithm runs in deterministic time $2^{O(n^2/m)}$ for any stretch $m\geq n+1$. In particular, this recovers the exponential-time brute force algorithm for the hardest case of $m=n+1$. It would be interesting to obtain matching conditional lower bounds for deterministic algorithms for $\NC^0_3$-\Avoid{}.
 \begin{openproblem}
     Is there a conditional lower bound of $2^{\Omega(n^2/m)}$ on the complexity of deterministic algorithms without an $\NP$ oracle for $\NC^0_3$-\Avoid{}?
\end{openproblem}

Finally, it is natural to ask if algorithms with $\NP$ oracles can solve $\NC^0_3$-\Avoid{} more efficiently. 
\begin{openproblem}
     Do there exist polynomial-time algorithms with $\NP$ oracles that solve $\NC^0_3$-\Avoid{} for stretch $m=o(n^2/\log(n))$?  
\end{openproblem}

\subsection{Structure}
The rest of the paper is organized as follows. In \cref{sec:prelims}, we give all necessary background material,
including a reduction from degree-$d$-\Avoid{} to $\NC^0_{d+1}$-\Avoid{} in \cref{sec:locality-reduction}. In \cref{sec:hardness}, we reduce the problem of constructing explicit rigid matrices to $\NC^0_3$-\Avoid{}. In \cref{sec:upper}, we give deterministic algorithms solving $\NC^0_k$-\Avoid{} in polynomial time for stretch $m\geq n^{k-1}/\log(n)$. Finally,
\cref{sec:appendix} contains an alternative deterministic polynomial-time algorithm for $\NC^0_3$ for the case where stretch $m\geq\Omega(n^2)$. 
\section{Preliminaries}\label{sec:prelims}

For every $a \in \mathbb{F}_2^n$ and subspace  $L$ of $\mathbb{F}_2^n$, we can define an affine subspace $\mathcal{A} \subseteq \mathbb{F}_2^n$ where $\mathcal{A} = \{a + v \mid v \in L\}$. The dimension of the affine subspace $\dim(\mathcal{A})$ is the same as the dimension of the linear subspace $L$ that defines it. Equivalently, the set of points that lie on a specified set of hyperplanes over $\mathbb{F}_2^n$ also characterize an affine subspace of $\mathbb{F}_2^n$. The hyperplanes can be written as a system of linear equations $Ax = b$, and the dimension of the corresponding affine subspace $\mathcal{A}$ can be calculated as $\dim(\mathcal{A}) = n - \rank(A)$. 

The circuits and algorithms in this paper generally work over the boolean hypercube $\{0, 1\}^n$. We work with multi-output circuits $\Ck: \{0, 1\}^n \rightarrow \{0, 1\}^m$ where $m > n$ and $m$ is called the stretch of the circuit. A partial output assignment, $y \in \{0, 1, \ast\}^m$, is a fixing of a subset of the output bits of the circuit to constants. For an input $x\in\{0,1\}^n$ to the circuit, we say \textit{$x$ is consistent with a partial output assignment $y \in \{0, 1, \ast\}^m$}, if $\Ck(x)$ agrees with $y$ on the fixed bits. 
When specified, the input (resp. output) space of a circuit might instead be viewed as the vector space $\mathbb{F}_2^n$ (resp. $\mathbb{F}_2^m$) over the finite field $\mathbb{F}_2$.

The complexity classes $\FP, \FP^\NP, \FE$, and $\FE^\NP$ are classes of search problems analogous to the classes of decision problems $\P, \P^{\NP}, \E$, and $\E^\NP$. For example, the class $\FP$ contains all functions that can be computed by deterministic polynomial-time Turing machines. 

\subsection{Circuits and Matrix Rigidity}
In this paper, we work with circuit classes $\NC^0_k$ and $\NC^1$, which we define below.
\begin{definition} \label{def:nc} \textbf{($\NC$ Circuits)}
The circuit class $\NC^i$ contains multi-output Boolean circuits on $n$ inputs of depth $O(\log^i(n))$ where each gate has fan-in $2$. We are particularly concerned with the following classes of circuits: 
\begin{itemize}
    \item For every constant $k\geq1$, $\NC_k^0$ is the class of circuits where each output depends on at most $k$ inputs.
    \item $\NC^1$ is the class of circuits of depth $O(\log(n))$ where all gates have fan-in $2$. 
    \item Linear $\NC^1$ circuits are circuits of depth $O(\log(n))$ where every gate has fan-in $2$ and computes an affine function, i.e., the XOR of its two inputs or its negation.
\end{itemize}
\end{definition}

It is a long-standing open problem in circuit complexity to prove super-linear lower bounds on the size of (linear) $\NC^1$ circuits computing an $n$-output function from $\FP$ or even $\FE^{\NP}$~\cite[Frontier~3]{valiant1977graph, AB2009}. 
Valiant~\cite{valiant1977graph} suggested an approach for proving super-linear lower bounds for linear $\NC^1$ circuits using the notion of matrix rigidity.

\begin{definition}[Matrix Rigidity] \label{def:row_rigid}
For $r, s \in \mathbb{Z}^{+}$, a matrix $M \in \mathbb{F}_2^{n \times n}$ is \emph{$(r, s)$-rigid} if $M$ cannot be written as a sum
\begin{align*}
    M = L + S \,,
\end{align*}
where $L, S \in  \mathbb{F}_2^{n \times n}$, $L$ is low rank, i.e., $\rank(L) \leq r$, and $S$ is row sparse, i.e., every row of $S$ has at most $s$ non-zero entries. 
\end{definition}

Valiant~\cite{valiant1977graph} proved that a linear operator given by a sufficiently rigid matrix requires linear $\NC^1$ circuits of size at least $\Omega(n\log\log(n))$, but there are still no known constructions of such rigid matrices even in $\FE^\NP$.

\begin{theorem}[{\cite{valiant1977graph}}]\label{thm:valiant}
    If a family of matrices $(M_n)_{n\geq1}$, $M_n \in \mathbb{F}^{n \times n}$, is $(\epsilon n, n^{\delta})$-rigid for constant $\epsilon, \delta > 0$, then the linear map $x \mapsto Mx$ requires linear $\NC^1$ circuits of size $\Omega(n \log \log (n))$.
\end{theorem}

\subsection{Range Avoidance for Circuits}
In the range avoidance problem, given a circuit $\Ck$ with $n$ inputs and $m$ outputs, $m>n$, the goal is to find an $m$-bit string outside the range of~$\Ck$.
\begin{definition}[\Avoid] \label{def:avoid}
In the \emph{\Avoid} problem, given a description of a circuit $\Ck : \{0, 1\}^n \rightarrow \{0, 1\}^{m}$ for $m > n$, the task is to find a $y \in \{0, 1\}^{m}$ such that $\forall x\in\{0,1\}^n\colon \Ck(x)\neq y$.
\end{definition}

The function $m=m(n)$ is called the \emph{stretch} of the multi-output circuit $\Ck$. Note that \Avoid{} is a total search problem, i.e., there always exists such a $y \in \{0, 1\}^{m}$ since $m > n$. 
We focus on a more restricted problem where there is an additional promise that the input circuit~$\Ck$ is from a fixed circuit class $\mathcal{C}$.
\begin{definition}[$\mathcal{C}$-\Avoid] \label{def:circ avoid}
In the \emph{$\mathcal{C}$-\Avoid} problem, given a description of a circuit $\Ck : \{0, 1\}^n \rightarrow \{0, 1\}^{m}$ for $m > n$, where $\Ck \in \mathcal{C}$, the task is to find a $y \in \{0, 1\}^{m}$ such that $\forall x\in\{0,1\}^n\colon \Ck(x)\neq y$.
\end{definition}

In particular, we are concerned with $\NC^1$-\Avoid{} and  $\NC_k^0$-\Avoid{} for constant $k \geq 1$. We will also consider the class of functions where each output is a multivariate polynomial of the inputs of degree at most $d$ over $\mathbb{F}_2$.
\begin{definition}[degree-$d$-\Avoid] \label{def:deg-d-avoid}
In the \emph{degree-$d$-\Avoid} problem, given a description of a function $\Ck\colon \F_2^n \to \F_2^{m}$ for $m > n$, where each output can be computed by a polynomial of degree $\leq d$ in the~$n$ inputs, the task is to find a $y \in \F_2^{m}$ such that $\forall x\in\F_2^n\colon \Ck(x)\neq y$.
\end{definition}

\subsection{Low Degree and Low Locality}\label{sec:locality-reduction}

\emph{Perfect randomized encodings} were introduced by \cite{applebaum2006cryptography} for various cryptographic applications. 
We are interested in the following property of perfect encodings: For a Boolean function $f:\{0,1\}^n \rightarrow \{0,1\}^{m}$ and its encoding $\widehat{f}\colon\{0,1\}^{n+\ell}  \to \{0,1\}^{m+\ell}$, there exists a polynomial-time decoding algorithm, $\mathrm{Dec}\colon \{0,1\}^{m+\ell} \to \{0,1\}^{m}$, such that for all $y \in \{0,1\}^{m+\ell}$ and  $x \in \{0,1\}^n$ satisfying $\mathrm{Dec}(y)= f(x)$, there exists $r \in \{0,1\}^{\ell}$ such that $y= \widehat{f}(x,r)$. This property can be used in \Avoid{} reductions as follows. Given a solution to the \Avoid{} problem for the function~$\widehat{f}$, i.e., $y\not\in\mathrm{Range}(\widehat{f})$, one can find a solution to the \Avoid{} problem for the function $f$ in polynomial time by simply computing $\mathrm{Dec}(y)\not\in\mathrm{Range}(f)$.

\cite{applebaum2006cryptography} first encode $\NC^1$ functions as degree-$3$ functions. Then, they encode every \mbox{degree-$d$} function as an $\NC_{d+1}^0$ function. Composing these two encodings provides an encoding of $\NC^1$ functions in $\NC_{4}^0$. Using this encoding, \cite{ren2022range} provides a polynomial time reduction from $\NC^1$-\Avoid{} to $\NC^0_4$-\Avoid{}. 
We use only one part of the result from~\cite{applebaum2006cryptography}: there is a polynomial time reduction from degree-$d$-\Avoid{} to $\NC_{d+1}^0$-\Avoid{}. For completeness, we include the proof here.

\begin{lemma}\label{lem:degree-encoding}
    Let $d\geq2$ be a constant, and $f\colon\F_2^n\to\F_2^m$ be a multi-output function where every output computes a sum of $k$ monomials of degree $\leq d$. 
    Then there exists a function $\widehat{f}\colon\{0,1\}^{n+(2k-1)m}\to\{0,1\}^{2km}$ computed by an $\NC^0_{d+1}$ circuit and a polynomial time algorithm $\mathrm{Dec}\colon\{0,1\}^{2km}\to\{0,1\}^{m}$ such that for all $x, y$, if $\mathrm{Dec}(y) = f(x)$, there exists an $r\in\{0,1\}^{(2k-1)m}$ such that $\widehat{f}(x, r) = y$.
    \begin{proof}
    We follow the encoding constructed in \cite{applebaum2006cryptography}. First, we construct an encoding $\widehat{g}$ for each single output function $g$ of $f$. 
      Let $g(x) =T_1(x)+ T_2(x)+\dots+ T_k(x)$ be a single output degree-$d$ function where each $T_i(x)$ is a monomial of degree at~most $d$. 
    Consider the encoding of $g$, $\widehat{g}:\{0,1\}^{n} \times \{0,1\}^{k} \times \{0,1\}^{k-1} \rightarrow \{0,1\}^{2k}$ defined as follows
     \begin{align*}
        \widehat{g}(x,r, s) = ( & T_1(x)-r_1, & & T_2(x)-r_2, & & \dotsc & & T_{k-1}(x)-r_{k-1}(x), & & T_k(x)-r_k,\\
            & r_1- s_1, & & s_1 + r_2 - s_2, & & \dotsc & & s_{k-2} + r_{k-1} - s_{k-1}, & & s_{k-1} + r_k) \,.
    \end{align*}
    Clearly, each output bit of $\widehat{g}$ can be computed by an $\NC^0_{d+1}$ circuit.
    We define a polynomial-time algorithm $\mathrm{Dec}_{\widehat{g}}\colon\{0,1\}^{2k}\to\{0,1\}$ such that if $\mathrm{Dec}_{\widehat{g}}(y)=g(x)$ then there exist $r$ and $s$ satisfying $\widehat{g}(x,r,s)=y$. 
      Given $y \in \{0,1\}^{2k}$, $\mathrm{Dec}_{\widehat{g}}(y)$ sums up the bits of $y$ modulo~$2$. 
      
      Suppose $\mathrm{Dec}_{\widehat{g}}(y)= g(x)$ for some $x \in \{0, 1\}^n$, i.e.,
      \begin{equation} \label{eqn:dec_single}
          \mathrm{Dec}_{\widehat{g}}(y) = \sum_{j=1}^{2k}y_j= g(x) = \sum_{j=1}^{k}T_{j}(x) \,.
      \end{equation}
      We will now show that there exist $r$ and $s$ such that $\widehat{g}(x,r,s)=y$.
      For each $j\in[k]$, we set $r_j = T_j(x) - y_j$.
      We also set $s_1 = r_1 - y_{k+1}$ and sequentially set $s_j = s_{j-1} + r_j - y_{k + j}$ for each $j \in \{2, \hdots, k-1\}$. By definition, the first $2k-1$ bits of $\widehat{g}(x, r, s)$ equal the first $2k-1$ bits of~$y$. For the last bit, note that:
      \begin{align*}
          s_{k-1} + r_k &= \sum_{i=1}^{k} r_i - \sum_{i=k+1}^{2k-1} y_i  & \text{(by the definition of $s$)} \\
          &= \sum_{i=1}^k T_i(x) - \sum_{i=1}^{2k-1} y_i & 
          \text{(by the definition of $r$)}\\
          &= y_{2k}\,. & \text{(by \cref{eqn:dec_single})}
      \end{align*}

      Therefore, for the constructed $r$ and $s$, $\widehat{g}(x, r, s) = y$, as required.

        Suppose $f(x)= (f_1(x), f_2(x), \dots, f_m(x))$, where each $f_i(x)$ is a sum of at most~$k$ monomials of degree $\leq d$. Let $\widehat{f}_i$ be the encoding of $f_i$ as defined above. Then our encoding of~$f$ is simply a concatenation of the encodings of its individual outputs,
         $\widehat{f}\colon \{0,1\}^{n+(2k-1)m} \to \{0, 1\}^{2km}$, where
        \begin{equation}
        \widehat{f}(x, r^{(1)}, r^{(2)}, \ldots, r^{(m)}, s^{(1)}, s^{(2)}, \ldots, s^{(m)}) = (\widehat{f}_1(x,r^{(1)},s^{(1)}),\ldots,\widehat{f}_m(x,r^{(m)},s^{(m)}))  \,.
        \end{equation}

        On input $y = (y_1,y_2,\ldots,y_m) \in \{0,1\}^{2km}$, the decoding algorithm returns 
        \begin{equation}
            \mathrm{Dec}(y)= (\mathrm{Dec}_{\widehat{f_1}}(y_1),\ldots, \mathrm{Dec}_{{\widehat{f_m}}}(y_{m})) \,.
        \end{equation}
        
        Suppose $\mathrm{Dec}(y)= f(x)$ for some $x \in \{0, 1\}^n$. Then $\mathrm{Dec}_{\widehat{f_i}}(y_i) = f_i(x)$ for all $i \in [m]$. By our proof above, there exists  $r^{(i)}$ and $s^{(i)}$ such that $y_i=\widehat{f}_i(x,r^{(i)},s^{(i)})$ and, thereby,
    \begin{align*}
        y= (y_1,\ldots,y_m)=(\widehat{f}_1(x,r^{(1)},s^{(1)}),\dots,\widehat{f}_m(x,r^{(m)},s^{(m)}))=\widehat{f}(x,r^{(1)},s^{(1)},\ldots,r^{(m)},s^{(m)})\,.
    \end{align*}

    Finally, $\mathrm{Dec}$ runs in time $O(mk)$ since it runs $m$ iterations of $\mathrm{Dec}_{\widehat{f_i}}$ for each $i$, each of which simply computes a sum of $2k$ bits. Since each $\widehat{f_i}$ is in $\NC^0_{d+1}$, so is $\widehat{f}$. 
    \end{proof}
\end{lemma}

Now, following~\cite{ren2022range}, we conclude that there is a polynomial-time reduction from degree-$d$-\Avoid{} to $\NC_{d+1}^0$-\Avoid{}. 
\begin{corollary} \label{cor:locality_reduction}
    For every $d\geq1$, if there exists an $\FP$ (resp. $\FP^\NP$) algorithm for $\NC^0_{d+1}$-\Avoid{}, then there exists an $\FP$ (resp. $\FP^\NP$) algorithm for degree-$d$-\Avoid{}. 
    \begin{proof}
        Let $f$ be an input to a degree-$d$-\Avoid{} problem with $m$ output bits. Then, each output bit of $f$ is a sum of at most $k=O(n^d)$ monomials of degree~$d$. Let $\widehat{f}$ be the encoding of $f$ in $\NC^0_{d+1}$ guaranteed by \cref{lem:degree-encoding}. Note that $\widehat{f}\colon \{0, 1\}^{n+(2k-1)m} \to \{0, 1\}^{2km}$. By the assumption of the Corollary, there is an $\FP$ (resp. $\FP^\NP$) algorithm that returns a $y \not \in \mathrm{Range}(\widehat{f})$. Then, by \cref{lem:degree-encoding}, $\mathrm{Dec}(y) \not \in \mathrm{Range}(f)$ and $\mathrm{Dec}$ runs in polynomial time. Therefore, there is an $\FP$ (resp. $\FP^\NP$) algorithm for degree-$d$-\Avoid{}.
    \end{proof}
\end{corollary}

\section{Hardness of \texorpdfstring{$\NC^0_3$-\Avoid{}}{NC0-Avoid}}\label{sec:hardness}
In this section, we reduce the problem of constructing explicit rigid matrices to the algorithmic task of solving $\NC^0_3$-\Avoid{}. First, in \cref{lem: rigid_sparse_lemma} we give an explicit degree-2 function $f\colon\{0,1\}^k\to\{0,1\}^n$, $k\ll n$, whose range contains all sparse vectors of length~$n$. Note that such a function $f$ must have degree at least~$2$. Indeed, if $f$ was affine and its range contained all vectors of sparsity at most~$1$, then its range must have dimension~$n$, and the number of inputs of $f$ would be $k\geq n$.

Next, in \cref{thm:rrlb}, we apply this lemma, together with the reduction from degree-$2$-\Avoid{} to $\NC^0_3$-\Avoid{} from \cref{lem:degree-encoding}, to conclude that an efficient algorithm for $\NC^0_3$-\Avoid{} would provide an explicit construction of rigid matrices.

\begin{lemma}\label{lem: rigid_sparse_lemma}
For every $d\geq1$ and every polynomial-time computable $s:=s(n)< \frac{n^{1 - 1/d}}{d}$, there exists a polynomial-time computable function  $f\colon \F_2^{dsn^{1/d}} \to \F_2^{n}$ whose range contains all vectors of sparsity at most $s$, and each output of $f$ is a degree-$d$ polynomial.
\end{lemma}

\begin{proof}
Let $G$ be an arbitrary $d$-uniform hypergraph on $\ell= d n^{1/d}$ vertices and $n$ hyperedges (such a graph exists because $\binom{\ell}{d}\geq (\ell/d)^d=n$). Fix an ordering $\{1, \hdots, \ell\}$ of the vertices and $\{1, \hdots, n\}$ of the edges. Each vertex of $G$ will be labeled by a vector from $\F_2^s$. Our function $f\colon\F_2^{s\ell}\to\F_2^{n}$ will take as input the labels of the vertices of $G$ and output $n$ elements corresponding to the $n$ hyperedges of $G$: the $i$th output is the generalized inner product of the labels of the $d$ vertices in the $i$th hyperedge. We interpret the input as a matrix $X \in \F_2^{s\times \ell}$, where the $j$th column $X_j \in \F_2^{s}$ is the label corresponding to the $j$th vertex. Suppose the hyperedge $i$ contains the vertices $\{j_1, \ldots, j_d\}$ then the $i$th output is $f_i(X) = \sum_{k=1}^s X_{k, j_1}\cdots X_{k, j_d}$. Clearly, $f$ is a degree-$d$ function, it only remains to show that its output contains all vectors of sparsity $\leq s$.
For this, we show that for every vector $y\in\F_2^n$ of sparsity $\leq s$, there is an input, i.e., a labeling of the vertices of~$G$, such that $f$ outputs $y$.
Let the $s$ non-zero elements of~$y$ correspond to the distinct edges $i_1, \hdots, i_s$ in~$G$. For each vertex $j$ in $G$ we set its label $X_j \in \F_2^s$ to be such that $(X_j)_k = 1$ if $j \in i_k$ and $(X_j)_k = 0$ otherwise.

Consider any edge $i = \{j_1, \hdots, j_d\}$ and the submatrix $X_{j_1, \hdots, j_d}$ of $X$ containing the labels of these vertices connected by $i$. 
\begin{itemize}
    \item If $i = i_k$ for some $k \in [s]$, then $y_i = 1$. The $i_k$th row of $X_{j_1, \hdots, j_d}$ contains all $1$ entries. Furthermore, every other row contains at least one zero. Therefore, $f_i(X) = 1$.
    \item If $i \neq i_k$ for all $k \in [s]$, then $y_i = 0$ and each row of $X_{j_1, \hdots, j_d}$ contains at least one zero. Therefore, $f_i(X) = 0$.\qedhere
\end{itemize}
\end{proof}

Equipped with \cref{lem: rigid_sparse_lemma}, we are ready to show that an efficient algorithm for degree-$2$-\Avoid{} or $\NC^0_3$-\Avoid{} would imply an explicit construction of rigid matrices.

\begin{theorem}\label{thm:rrlb}
    For every constant $1/2\leq \delta \leq 1$, an $\FP$ (resp. $\FP^\NP$) algorithm for degree-$2$-\Avoid{} with stretch $m = 2n^{2/(1+\delta)}$ will provide an $\FP$ (resp. $\FP^\NP$) algorithm for finding an $(n^{\delta}/10, n^{\delta-1/2}/10)$-rigid matrix. 
    
    Furthermore, for every $1/2\leq \delta \leq 1$, an $\FP$ (resp. $\FP^\NP$) algorithm for $\NC^0_3$-\Avoid{} with stretch $m=n+O(n^{2/(2+\delta)})$ will provide an $\FP$ (resp. $\FP^\NP$) algorithm for finding an $(n^{\delta}/10, n^{\delta-1/2}/10)$-rigid matrix.
\end{theorem}

\begin{proof}
Let $r=n^{\delta}/10$ and $s=n^{\delta-1/2}/10$. First, we reduce (in deterministic polynomial time) the problem of finding an $(r, s)$-rigid matrix to solving degree-$2$-\Avoid{} for a function $g\colon \F_2^{4rn} \to \F_2^{n^2}$.

Suppose $M \in \F_2^{n \times n}$ is not $(r, s)$-rigid. Then, $M$ can be written as a sum $M = Q + S$, where $\rank(Q) \leq r$ and $S$ is $s$-row sparse. Furthermore, $Q = L \cdot R$ for some matrices $L,R^T \in \mathbb{F}_2^{n \times r}$. 

We view the input of $g$ as $2rn$ entries of the matrices $L$ and $R$, and $2sn^{3/2}$ inputs of $n$ copies of the degree-$2$ function~$f$ from \cref{lem: rigid_sparse_lemma} needed to encode the entries of $n$ sparse rows of~$S$. Then the function $g$ simply outputs all $n^2$ entries of $M=L\cdot R + S$. Note that each output of $g$ computes a dot-product of a row of $L$ and a column of $R$, and adds a degree-$2$ output of $f$. Therefore, we constructed a degree-$2$ function $g$ whose range contains all non-rigid matrices. A~solution to degree-$2$-\Avoid{} on input $g$ would therefore give an $(r, s)$-rigid matrix. The number of inputs of~$g$ is
$n'=2 rn + 2sn^{3/2}=4rn=2n^{1+\delta}/5$, and the stretch of the function $g$ is at least $m'(n')\geq 2(n')^{2/(1+\delta)}$. This concludes the proof of the first part of the theorem.

For the second part, we use the polynomial-time reduction from degree-$2$-\Avoid{} to $\NC^0_3$-\Avoid{} from \cref{lem:degree-encoding}. By the construction above, we have a degree-$2$ function $g\colon \F_2^{4rn} \to \F_2^{n^2}$ where each output bit is the sum of at most $t = r+s\leq 2r$ degree-$2$ monomials.  We apply \cref{lem:degree-encoding} to reduce \Avoid{} for $g$ to \Avoid{} for an $\NC^0_3$ function $\widehat{g}\colon \{0,1\} ^{\widehat{n}}   \to \{0,1\}^{\widehat{m}}$, where $\widehat{n}=n' + (2t - 1)n^2$ and $\widehat{m}=2tn^2$. This yields a stretch of $\widehat{m}(\widehat{n}) = \widehat{n}+O(\widehat{n}^{2/(2+\delta)})$ for the function $\widehat{g}$. Therefore, an algorithm for $\NC^0_3$-\Avoid{} for stretch $\widehat{m}(n)$ yields an $(r, s)$-rigid matrix. 
\end{proof}
We remark that in the regime $\delta>1/2$, \cref{thm:rrlb} would give matrices that for rank~$n^{\delta}$ have higher rigidity  than all known constructions of rigid matrices in $\FP, \FP^{\NP}$ and $\FE^{\NP}$. Therefore, for every $\eps>0$, an $\FP^\NP$ algorithm for degree-$2$-\Avoid{} with stretch $n^{12/11-\eps}$ or an $\FP^\NP$ algorithm for $\NC^0_3$-\Avoid{} with stretch $n+n^{12/17+\eps}$ would lead to new rigidity lower bounds. Since the regime of $\delta=1$ in \cref{thm:rrlb} is sufficient for Valiant's program of proving super-linear lower bounds on the size of linear $\NC^1$ circuits (see \cref{thm:valiant}), we have the following corollary. 

\begin{corollary}\label{cor:clb}
 An $\FP$ (resp. $\FP^\NP$) algorithm for degree-$2$-\Avoid{} with stretch $m= 2n$ or for $\NC^0_3$-\Avoid{} with stretch $m=n+O(n^{2/3})$ will provide a linear function in $\FP$ (resp. $\FP^\NP$) that cannot be computed by linear $\NC^1$ circuits of size $o(n\log\log(n))$.
\end{corollary}

\section{Algorithms for \texorpdfstring{$\NC^0_k$-\Avoid{}}{NC0-Avoid}}\label{sec:upper}

In this section, we describe polynomial-time algorithms for solving $\NC_k^0$-\Avoid{} with non-trivial stretch. More specifically, we provide an algorithm that runs in time $2^{O(n^{k-1}/m)}\cdot\poly(n)$ when the stretch of the input circuit is at least $m \geq \Omega(n^{k-2})$. First, we describe a useful structural property of $\NC_k^0$ circuits, which follows from the following simple graph-theoretic result. 

\begin{lemma} \label{lem:hypergraph}
For any constants $c \geq 1$ and $k\geq 3$, every $k$-uniform hypergraph $G = (V, E)$ with $n$ vertices and $m\geq cn^{k-2}$ hyperedges contains a subset of vertices $V'\subseteq V, |V'| \leq cn^{k-1}/m$ and a subset of hyperedges $E'\subseteq E,|E'| \geq cn$ such that each hyperedge in $E'$ contains at least $k-2$ vertices from~$V'$. Furthermore, there is a polynomial-time algorithm that finds such a $V'$ and $E'$.
\begin{proof}
Consider the following bipartite graph $H$ with vertex set $A\sqcup B$ where $A = \{u_S \mid S \subseteq V, |S| = k - 2\}$ is the set of vertices indexed by the $k-2$ sized subsets of $V$ and $B = \{v_e \mid e \in E\}$ is indexed by the edges of $G$. Furthermore there is an edge $(u_S,  v_e)$ in $H$ if $S \subseteq e$, i.e., if all the vertices in $S$ are contained in the hyperedge $e \in E$. Since each hyperedge $e$ contains $k$ vertices, the degree of each vertex in $B$ is exactly $\binom{k }{k - 2}$. Then, the average degree of the vertices in $A$ is $\frac{|B| \binom{k }{k - 2}}{|A|} = \frac{m \binom{k}{k - 2}}{\binom{n }{k - 2}}$. Let $A' \subseteq A$ be the subset of $t$ vertices with highest degree in $A$ and let $N(A')$ be their neighbors in $B$. Then total degree of vertices in $A'$ is at least $\frac{tm \binom{k }{k - 2}}{\binom{n }{k - 2}}$. Since each vertex in $B$ has degree $\binom{k }{ k - 2}$, $|N(A')| \geq \frac{tm \binom{k }{ k - 2}}{\binom{n }{k - 2}{\binom{k }{k - 2}}} = \frac{tm}{\binom{n }{k - 2}}$.  Therefore, setting $t = \frac{cn\binom{n }{ k - 2}}{m}$, $V'$ to be the set of $t(k-2) = \frac{cn(k-2)\binom{n }{k -2}}{m}\leq cn^{k-1}/m$ vertices of $G$ contained in the union of the vertex subsets in $A'$, and $E' = N(A')$ completes our proof. 

To find $V'$ and $E'$, we first construct the graph~$H$ which has polynomial size, and then find the vertices in $A'$ by finding the $t$ vertices in $A$ with maximum degree. It is now straightforward to construct $V'$ from $A'$ and to find $E' = N(A')$.  
\end{proof}
\end{lemma}

\begin{corollary} 
\label{cor:dense_inp_out_map}
For any constants $c \geq 1$ and $k\geq 3$, given an $\NC^0_k$ circuit $\Ck$ with $n$ inputs and $m \geq cn^{k-2}$ outputs, there exists a subset of outputs $O$ of size $|O|\geq cn$, and a subset of inputs $I$ of size $|I|\leq cn^{k-1}/m$, such that for every output bit $\Ck_i \in O$, at least $k-2$ of the input bits feeding into $\Ck_i$ are from $I$. Furthermore, there is a polynomial-time algorithm that finds such sets $I$ and $O$.

\begin{proof}
    Without loss of generality we assume that each output of $\Ck$ reads exactly $k$ inputs (as if it reads $\ell<k$ inputs, we let it additionally read arbitrary $k-\ell$ inputs and ignore them).
    Consider the hypergraph where each vertex corresponds to one of the $n$ inputs $\{x_1,\ldots,x_n\}$ of $\Ck$. Each edge of the hypergraph corresponds to an output $\Ck_i$, $e_i = \{j \mid \Ck_i \text{ reads } x_j\}$. Now, we apply \cref{lem:hypergraph} on this hypergraph and set $I = V'$ and $O = E'$. Note that $|I| = |V'| \leq cn^{k-1}/m$ and $|O| = |E'| \geq cn$. 
\end{proof}
\end{corollary}

This corollary finds a linear number of output bits $O$ of the circuit that mostly depend on a small number of common input bits~$I$.  Our algorithm for $\NC^0_k$-\Avoid{} will ``branch'' on all possible assignments to the inputs from~$I$. Each such assignment will correspond to an affine subspace $\AS\subseteq\F_2^n$ of the input space.  Then, our algorithm works by fixing the output bits from~$O$ such that the sum of the dimensions of these affine subspaces is significantly reduced at each step, eventually making all subspaces empty. Note that by the guarantee of \cref{cor:dense_inp_out_map}, after fixing the inputs~$I$, each output from~$O$ depends on at most two inputs. Thus, we need an efficient way to reduce the dimension of the affine subspace containing the consistent inputs for the case where output functions depend on at most two inputs. In \cref{lem:nc02-avoid}, we provide such a subroutine $\textsc{AffineReduce}$ (\cref{alg:dimension}).

\begin{lemma}\label{lem:nc02-avoid}
Let $\AS\subseteq\F_2^n$ be an affine subspace, and $f\colon\F_2^n\to\F_2$ be a function that depends on at most two inputs. The algorithm $\textsc{AffineReduce}$ in deterministic polynomial time finds two affine subspaces (or empty sets) $\AS_0,\AS_1\subseteq \AS$ such that
\begin{enumerate}
    \item [(1)] $\forall x\in \AS, b\in\F_2$, if $f(x)=b$, then $x\in\AS_b$;
    \item [(2)] $|\AS_0| + |\AS_1| \leq 3|\AS|/2$.
\end{enumerate}
\end{lemma}

\begin{proof}
Without loss of generality we assume that $f(x)$ depends on (a subset of) $x_1$ and $x_2$. We will consider three cases depending on the degree of~$f$, and in each case we will find affine subspaces (or empty sets) $\AS_0,\AS_1\subseteq \AS$ such that at least one of them has dimension strictly smaller than the dimension of $\AS$ (or at least one of them is an empty set). This will ensure that $|\AS_0| + |\AS_1| \leq 3|\AS|/2$.
\begin{itemize}
    \item If $f(x)=c$ for some $c\in\F_2$ is a constant function, then we set $\AS_c=\AS$ and $\AS_{1-c}=\emptyset$. Clearly, $\AS_c = \AS$ and $\AS_{1-c}$ contain all points $x\in\AS$ that are consistent with $f(x)=c$ and $f(x)=1-c$, respectively.
    \item If $f(x) = a_1x_1 + a_2x_2 + c$ for some constants $a_1, a_2, c \in \F_2$ is an affine function, then for each $b\in\F_2$ let $H_b$ be the hyperplane defined by $a_1x_1 + a_2x_2 + c = b$, and let $\AS_b = \AS \cap H_b$. Again, $\AS \cap H_b$ contains all the inputs in $\AS$ that are consistent with $f(x)=b$. Furthermore, if $\dim(\AS_0)=\dim(\AS)$, then $\AS\subseteq H_0$ and $\AS_1= \AS \cap H_1=\emptyset$. Therefore, either $\dim(\AS_0)<\dim(\AS)$ or $\AS_1=\emptyset$.
    \item If $f(x) = (x_1 + a_1)(x_2 + a_2) + c$ for some constants $a_1, a_2, c \in \F_2$ is a quadratic function, then let $\mathcal{H}$ be the affine subspace defined by $\mathcal{H}= \{x \in \F_2^n \mid x_1 = 1 + a_1, x_2 = 1 + a_2\}$. Consider the affine subspace $\AS_{1-c} = \AS \cap \mathcal{H}$ which contains all points $x\in\AS$ satisfying $f(x)=1-c$.
    \begin{itemize}
        \item If $\dim(\AS_{1-c}) < \dim(\AS)$, then we are done as we can take $\AS_{c} = S$ 
        \item If $\dim(\AS_{1-c}) = \dim(\AS)$, then $\AS\subseteq \mathcal{H}$. Then, every point in $\AS$ satisfies $f(x) = 1-c$, thus, setting $\AS_{1-c}=\AS$ and $\AS_c=\emptyset$ completes our construction.
    \end{itemize}
\end{itemize}
In each case, either $|\AS_0| \leq \frac{|\AS|}{2}$ or $|\AS_1| \leq \frac{|\AS|}{2}$. Therefore, 
$|\AS_0| + |\AS_1| \leq 3|\AS|/2$.

The only computation made by \textsc{AffineReduce} is to compute the dimensions of explicitly given affine subspaces, which can be performed in polynomial time.
\end{proof}

\begin{algorithm}[!ht]
\caption{$\textsc{AffineReduce}(\AS, f)$}\label{alg:dimension}
\begin{algorithmic}
\Require Affine subspace $\AS \subseteq \F_2^n$, $f\colon\F_2^n\to\F_2$ that may depend only on $x_1$ and $ x_2$
\Ensure $\AS_0, \AS_1 \subseteq \AS$
\If{$f(x) = c$}
    \State \Return $\AS_{c} = \AS$ and $\AS_{1-c}=\emptyset$
\EndIf
\If{$f(x) = a_1 x_1 + a_2 x_2 + c$}
    \State For $b\in\F_2$, let $H_b=\{x\in\F_2^n\colon a_1x_1 + a_2x_2 + c = b\}$ 
    \State \Return $\AS_0 = \AS \cap {H}_0$ and $\AS_1 = \AS \cap {H}_1$
\EndIf
\If{$f(x) = (x_1 + a_1)(x_2 + a_2) + c$}
    \State Let $\mathcal{H}= \{x\in\F_2^n\colon x_1 = 1 + a_1, x_2 = 1 + a_2\}$
    \State Let $\AS_{1-c} = \AS \cap \mathcal{H}$
    \If {$\dim(\AS_{1-c}) < \dim(\AS)$}
        \State \Return $\AS_{1-c}$ and $\AS_c=\AS$
    \Else
        \State  \Return $\AS_{1-c}$ and $\AS_c=\emptyset$
    \EndIf
\EndIf
\end{algorithmic}
\end{algorithm}

A simple application of \textsc{AffineReduce} recovers a polynomial-time algorithm for $\NC_2^0$-\Avoid{} from \cite{guruswami2022range}.

\begin{corollary}
There is a deterministic polynomial-time algorithm that, given an $\NC^0_2$ circuit $\Ck\colon \{0,1\}^n \to \{0,1\}^m$, $m \geq n+1$, finds an element $y \in \{0, 1\}^{m}, y \not \in \mathrm{Range}(\Ck)$.
\end{corollary}
\begin{proof}
 At iteration $1\leq i\leq n+1$, our algorithm will fix the value of the $i$th output bit $y_i$. The algorithm also maintains an affine subspace $\AS\subseteq\F_2^n$ that contains all inputs $x\in\F_2^n$ consistent with the partial output assignments of $y_1, \hdots, y_i$. By \cref{lem:nc02-avoid}, there exists an assignment $y_i=b$, such that either none of the inputs in $\AS_b$ are consistent with $y$ or the dimension of $\AS = \AS_b$ reduces at least by one. In the former case, we already find our desired output $y$ (we can just set the unassigned bits of $y$ to arbitrary values). Otherwise, after fixing the first $n$ outputs, we have $\dim(\AS)=0$, i.e., $S=\{x\}$ for some $x\in \F_2^n$. Let $b\in\{0,1\}$ be the value of the $(n+1)$th output bit of $\Ck(x)$. Then setting $y_{n+1} = 1-b$ produces our desired output $y$. 
This algorithm runs in polynomial time since it makes at most $n$ calls to $\textsc{AffineReduce}$ and one call to $\Ck(x)$.
\end{proof}
Finally, equipped with \cref{lem:nc02-avoid}, we are ready to present our main algorithm for $\NC_k^0$-\Avoid{}.

\begin{theorem} \label{thm:good-stretch}
Given an $\NC^0_k$ circuit $\Ck\colon \{0,1\}^n \to \{0,1\}^m$, where $m \geq 3n^{k-2}$, the algorithm \textsc{SubspaceUnion} finds an element $y \in \{0, 1\}^{m}, y \not \in \mathrm{Range}(\Ck)$ in deterministic time ${2^{O(n^{k-1}/m)}\cdot \poly(n)}$.
\end{theorem}

\begin{proof}
First we apply \cref{cor:dense_inp_out_map} with $c=3$ to the circuit $\Ck$, and select in polynomial time a subset of inputs $I=\{x_1,\ldots,x_t\}$ and a set of outputs $O=\{y_1,y_2,...., y_{3n}\}$ for  $t\leq 3n^{k-1}/m$. This ensures that each $y_i$ has at most two inputs outside of~$I$. For each of the $2^t$ assignments of the inputs from $I$, we consider a circuit where the values of these $t$ inputs are fixed.  Namely, for $j \in\{0, \ldots, 2^t-1\}$, we fix the inputs in~$I$ to the bits in the binary representation of $j$. Then we restrict the circuit $\Ck$ to the outputs $y_1, \hdots, y_{3n}$ and all the inputs that feed them, and obtain a circuit $\Ck_{j}$, where each output depends on at most $2$ inputs. We'll find a value $y\in\{0,1\}^{3n}$ that no $\Ck_{j}$ outputs, and this will give us a solution to the original $\NC^0_k$-\Avoid{} instance. 

Our algorithm will maintain the following invariant. At the $i$th iteration of the algorithm after we fix the values of the outputs $y_1,\ldots,y_i$, we maintain $\AU = \bigcup_{j=0}^{2^t-1}\AU_{j}$, a disjoint union of $2^t$ affine subspaces, such that all inputs $x\in\F_2^n$ that are consistent with $y_1,\ldots,y_i$ belong to $\AU$ (and $\AU$ may contain points that are inconsistent with $y_1,\ldots,y_i$, too). 

In the beginning of the algorithm, for every $0\leq j<2^t$, we let $\AU_{j}$ be the affine subspace where the inputs in~$I$ are fixed to the bits in the binary representation of $j$. Then $\AU = \bigcup_{j}\AU_{j}=\F_2^n$ is the set of all inputs consistent with our initial empty partial assignment.

 At every step $i$, we will show how to find a constant $b\in\{0,1\}$ such that after fixing $y_i = b$, the size of our disjoint union $|\AU|$ reduces by a factor of $4/3$. Therefore, after repeating this procedure for the $3n$ outputs from~$O$, we will have an empty $\AU$, and the constructed partial assignment will give us a solution to the $\NC^0_k$-\Avoid{} problem.

At the $i$th iteration of the algorithm, we have values of outputs $y_1, \hdots y_{i-1}$ fixed, and are to fix the value of $y_i$. We have two choices: either set $y_i = 0$ or set $y_i = 1$. By \cref{lem:nc02-avoid}, we have two affine subspaces (or empty sets) $\AU_{j,0},\AU_{j, 1}\subseteq\AU_j$ containing all inputs $x\in\AU_j$ mapping to $y_i=0$ and $y_i=1$, respectively. Moreover,  \cref{lem:nc02-avoid} guarantees that $|\AU_{j, 0}| + |\AU_{j,1}| \leq 3|\AU_{j}|/2$. Summing over all $0\leq j<2^t$, we get 
\[\sum_{j} |\AU_{j,0}| + \sum_{j}|\AU_{j,1}| \leq \sum_{j} 3|\AU_{j}|/2  = 3|\AU|/2\,.\] 
Let $b\in\{0,1\}$ be the value minimizing $\sum_{j} |\AU_{j,b}|$. In particular, we have that $\sum_{j} |\AU_{j,b}|\leq 3|\AU|/4$. Therefore, setting $y_i = b$ reduces the size of $\AU$ at least by a factor of $4/3$. Repeating this procedure $\log_{4/3}(2^n)+1 \leq 3n$ times will result in a partial assignment to the output bits $O$ with no inputs that map to it.

The algorithm \textsc{SubspaceUnion} maintains $2^t$ affine subspaces of $\F_2^n$, computes their dimensions and calls the deterministic polynomial-time $\textsc{AffineReduce}$ procedure polynomial number of times. Therefore, this algorithm runs in time $2^t\cdot\poly(n)=2^{O(n^{k-1}/m)}\cdot \poly(n)$.
\end{proof}

\begin{algorithm}[!ht]
\caption{$\textsc{SubspaceUnion}(\Ck)$}\label{alg:good_alg}
\begin{algorithmic}
\Require $\NC^0_k$ circuit $\Ck: \{0,1\}^n \rightarrow \{0,1\}^m$, where $m\geq 3n^{k-2}$
\Ensure $y \in \{0,1\}^{m}$, $y \notin \mathrm{Range}(\Ck)$

\State Find $x_1,\hdots,x_t$ and $y_1,\hdots,y_{3n}$ via \cref{cor:dense_inp_out_map} for $t\leq 3n^{k-1}/m$
\State For $0\leq j<2^t$, set $\AU_{j} = \{x\in\{0,1\}^n\colon \sum_{i=1}^{t} x_i2^{i-1}=j\}$
\For{i=1 to 3n}
        \State Find function $f$ at $y_i$
        \State For $0\leq j<2^t$, set $\AU_{j,0},\AU_{j,1}\gets\textsc{AffineReduce}(\AU_j,f)$
        \State Find $b\in\{0,1\}$ minimizing  $\sum_{j} |\AU_{j,b}|$
        \State Set $y_i=b$
        \State For $0\leq j<2^t$, set $\AU_{j}=\AU_{j,b}$

\EndFor
\State Set all remaining $y_k = 0$
\State \Return $y$
\end{algorithmic}
\end{algorithm}
We conclude this section with a corollary stating that \textsc{SubspaceUnion} solves $\NC^0_k$-\Avoid{} efficiently for certain non-trivial values of stretch~$m$.

\begin{corollary}
    For any constants $k\geq 3$ and $\eps>0$, the algorithm \textsc{SubspaceUnion} solves $\NC^0_k$-\Avoid{} on $n$ inputs and $m$ outputs in deterministic polynomial and deterministic sub-exponential $2^{O(n^{1-\eps})}$ time for $m\geq n^{k-1}/\log(n)$ and $m\geq n^{k-2+\eps}$, respectively.
\end{corollary}

\subsection*{Acknowledgements}

We thank Justin Thaler, Sam King, and anonymous reviewers for their helpful comments on our paper. 
\bibliographystyle{alpha}
\bibliography{nc0_avoid}

\begin{thebibliography}{KKMP21}

\bibitem[AB09]{AB2009}
Sanjeev Arora and Boaz Barak.
\newblock {\em Computational complexity: a modern approach}.
\newblock Cambridge University Press, 2009.

\bibitem[AC19]{ac19}
Josh Alman and Lijie Chen.
\newblock Efficient construction of rigid matrices using an {NP} oracle.
\newblock In {\em FOCS}, 2019.

\bibitem[AIK06]{applebaum2006cryptography}
Benny Applebaum, Yuval Ishai, and Eyal Kushilevitz.
\newblock Cryptography in $\textrm{NC}^0$.
\newblock {\em SIAM Journal on Computing}, 36(4):845--888, 2006.

\bibitem[BHPT20]{bhpt20}
Amey Bhangale, Prahladh Harsha, Orr Paradise, and Avishay Tal.
\newblock Rigid matrices from rectangular {PCPs}.
\newblock In {\em FOCS}, 2020.

\bibitem[Cha94]{chashkin}
Aleksandr~V. Chashkin.
\newblock On the complexity of {B}oolean matrices, graphs and their
  corresponding {B}oolean functions.
\newblock {\em Discrete Mathematics and Applications}, 4(3):229--257, 1994.

\bibitem[CHLR23]{CHLR23}
Yeyuan Chen, Yizhi Huang, Jiatu Li, and Hanlin Ren.
\newblock Range avoidance, remote point, and hard partial truth table via
  satisfying-pairs algorithms.
\newblock In {\em STOC}, 2023.

\bibitem[Fri93]{f93}
Joel Friedman.
\newblock A note on matrix rigidity.
\newblock {\em Combinatorica}, 13(2):235--239, 1993.

\bibitem[GLW22]{guruswami2022range}
Venkatesan Guruswami, Xin Lyu, and Xiuhan Wang.
\newblock Range avoidance for low-depth circuits and connections to
  pseudorandomness.
\newblock In {\em RANDOM}, 2022.

\bibitem[GT16]{goldreich2016matrix}
Oded Goldreich and Avishay Tal.
\newblock Matrix rigidity of random {Toeplitz} matrices.
\newblock In {\em STOC}, 2016.

\bibitem[IK00]{IK00}
Yuval Ishai and Eyal Kushilevitz.
\newblock Randomizing polynomials: A new representation with applications to
  round-efficient secure computation.
\newblock In {\em FOCS}, 2000.

\bibitem[IK02]{IK02}
Yuval Ishai and Eyal Kushilevitz.
\newblock Perfect constant-round secure computation via perfect randomizing
  polynomials.
\newblock In {\em ICALP}, 2002.

\bibitem[ILW23]{ILW23}
Rahul Ilango, Jiatu Li, and Ryan Williams.
\newblock Indistinguishability obfuscation, range avoidance, and bounded
  arithmetic.
\newblock In {\em STOC}, 2023.

\bibitem[IP99]{IP99}
Russell Impagliazzo and Ramamohan Paturi.
\newblock The complexity of {$k$-SAT}.
\newblock In {\em CCC}, 1999.

\bibitem[IPZ98]{IPZ98}
Russell Impagliazzo, Ramamohan Paturi, and Francis Zane.
\newblock Which problems have strongly exponential complexity?
\newblock In {\em FOCS}, 1998.

\bibitem[KKMP21]{kleinberg2021total}
Robert Kleinberg, Oliver Korten, Daniel Mitropolsky, and Christos
  Papadimitriou.
\newblock Total functions in the polynomial hierarchy.
\newblock In {\em ITCS}, 2021.

\bibitem[Kor21]{korten2022hardest}
Oliver Korten.
\newblock The hardest explicit construction.
\newblock In {\em FOCS}, 2021.

\bibitem[LY22]{LY22}
Jiatu Li and Tianqi Yang.
\newblock $3.1n-o(n)$ circuit lower bounds for explicit functions.
\newblock In {\em STOC}, 2022.

\bibitem[PR94]{pr94}
Pavel Pudl{\'a}k and Vojtech R{\"o}dl.
\newblock Some combinatorial-algebraic problems from complexity theory.
\newblock {\em Discrete Mathematics}, 1(136):253--279, 1994.

\bibitem[RSW22]{ren2022range}
Hanlin Ren, Rahul Santhanam, and Zhikun Wang.
\newblock On the range avoidance problem for circuits.
\newblock In {\em FOCS}, 2022.

\bibitem[Sha49]{Shannon49}
Claude~E. Shannon.
\newblock The synthesis of two-terminal switching circuits.
\newblock {\em The Bell System Technical Journal}, 28:59--98, 1949.

\bibitem[SSS97]{sss97}
Mohammad~A. Shokrollahi, Daniel~A. Spielman, and Volker Stemann.
\newblock A remark on matrix rigidity.
\newblock {\em Information Processing Letters}, 64(6):283--285, 1997.

\bibitem[Val77]{valiant1977graph}
Leslie~G. Valiant.
\newblock Graph-theoretic arguments in low-level complexity.
\newblock In {\em MFCS}, 1977.

\end{thebibliography}
\newpage
\appendix
\section{An Alternative Algorithm for \texorpdfstring{$\NC^0_3$-\Avoid{}}{NC0-Avoid}}\label{sec:appendix}

\begin{theorem}\label{thm:apx}
Given an $\NC^0_3$ circuit $\Ck: \{0,1\}^n \rightarrow \{0,1\}^{m}$, where $m \geq \frac{1}{3}\binom{n}{2} + 2n$, the algorithm \textsc{OneSubspace} finds an element $y \in \{0, 1\}^{m}, y \not \in \mathrm{Range}(\Ck)$ in deterministic polynomial time.

\begin{proof}

The algorithm maintains an affine subspace $\AS \subseteq \F_2^n$ over the inputs, and a partial output assignment $y \in \{0, 1, \ast\}^m$ such that $\AS$ contains all inputs $x\in\F_2^n$ consistent with $y$. Initially, $y = (\ast,\hdots,\ast)$ and $\AS = \F_2^n$. At each iteration, \textsc{OneSubspace} assigns at most two outputs and reduces the dimension of $\AS$ by at least~$1$. After $n$ steps, $\AS$ must have dimension~$0$. Then the algorithm assigns one more output bit, and terminates with an element $y \not \in \mathrm{Range}(\Ck)$.

Now, we only need to argue that the algorithm can reduce the dimension of $\AS$ in each iteration and that we can perform each step in polynomial time. 

First, if there is an output $y_{1}$ that depends on at most $2$ inputs $x_{1}, x_{2}$, let $f$ be the function computed at that output: $y_1=f(x_1, x_2)$. By \cref{lem:nc02-avoid}, \textsc{AffineReduce}$(\AS, f)$ outputs an affine subspace $\AS_b$ of lower dimension $\dim(\AS_b)<\dim(\AS)$, containing all inputs consistent with $y_1 = b$. Thus, in the following we assume that each output depends on exactly~$3$ inputs.

 Since we fix at most $2$ bits of the output at each iteration, the number of unassigned outputs $m$ is always greater than $\frac{1}{3}\binom{n}{2}$. 
  Then, there exists a pair of outputs $y_1, y_2$ that both depend on the same pair of inputs $x_2, x_3$. Since each output depends on three pairs of inputs, and the number of such pairs is $\binom{n}{2}<3m$, there must be a pair of inputs that feeds into two outputs. 
    Let $x_1, x_4$ be the remaining inputs that feed into the outputs $y_1, y_2$, respectively: $y_1 = f_1(x_1, x_2, x_3), y_2 = f_2(x_2, x_3, x_4)$. 
   
    There exist $4$ possible values of $(y_1, y_2)$ and at most $16$ possible values of the input bits $(x_1, x_2, x_3, x_4)$ appearing in~$\AS$. Then, there exists a pair of constants $(b_1, b_2)\in\{0,1\}^2$ such that at most $4$ different assignments $A\subseteq\{0,1\}^4$ to $(x_1, x_2, x_3, x_4)$ are consistent with the partial assignment $(y_1,y_2)=(b_1,b_2)$. Since $|A|\leq 4$, there is a $3$-dimensional affine subspace in $\F_2^4$ that contains all points from~$A$. Therefore, there is a hyperplane $\AH\subseteq\F_2^4$ defining this $3$-dimensional affine subspace. Extending $\AH$ to all $n$ inputs, gives us an affine subspace $\AH'=\{x\in\F_2^n\colon (x_1,x_2,x_3,x_4)\in\AH\}\subseteq\F_2^n$ that contains all inputs in $\F_2^n$ consistent with the partial assignment $(y_1,y_2)=(b_1,b_2)$. 
    
    If $\AS \not\subseteq \AH'$, then setting $(y_1, y_2) = (b_1, b_2), \AS = \AS \cap \AH'$ reduces the dimension of the affine subspace~$\AS$.
    In the following we assume that $\AS \subseteq \AH'$. 
        \begin{itemize}
            \item Suppose there exists $(c_1, c_2)\in\{0,1\}^2$ such that for all points $(x_1, x_2, x_3, x_4)$ in $\AH'$, $(f_1(x_1, x_2, x_3), f_2(x_2, x_3, x_4)) \neq (c_1, c_2)$. Then we can set $(y_1, y_2) = (c_1, c_2)$ and $\AS = \emptyset$ as no points in $\AS \subseteq \AH'$ can output $(c_1, c_2)$. In this case we found a $y\not\in\Range(\Ck)$.
            \item If there are no such assignments, then since $|\AH| =8$, there must exist an assignment $(c_1, c_2)\in\{0,1\}^2$ such that at most two points from $\AH$ are consistent with $(y_1,y_2)=(c_1, c_2)$. These (at most) two points form a $0$- or $1$-dimensional affine subspace $\AU\subseteq\F_2^4$, which we extend to all $n$ inputs $\AU'=\{x\in\F_2^n\colon(x_1,x_2,x_3,x_4)\in\AU\}\subseteq\F_2^n$.
            \begin{itemize}
                \item If $\AS \not \subseteq \AU'$, we can set $(y_1, y_2) = (c_1, c_2)$ and $\AS = \AS \cap \AU'$, reducing the dimension of $\AS$.
                \item Otherwise, all inputs in $\AS\subseteq\AU'$ have $(y_1,y_2)=(c_1, c_2)$, and we can set $(y_1, y_2) = (1-c_1, c_2)$ to obtain $\AS = \emptyset$. 
            \end{itemize}
        \end{itemize}
     
This algorithm performs $n$ iterations, each of which computes dimensions of a constant number of explicitly given affine subspaces in polynomial time.
\end{proof}
\end{theorem}
\begin{algorithm}[!ht]
\caption{$\textsc{OneSubspace}(\Ck)$}\label{alg:bad_alg}
\begin{algorithmic}
\Require $\NC^0_3$ circuit $\Ck: \{0,1\}^n \rightarrow \{0,1\}^m$, where $m\geq \frac{1}{3}\binom{n}{2}+2n$
\Ensure $y \in \{0,1\}^{m}$, $y \notin \mathrm{Range}(\Ck)$
\State Let $\AS = \F_2^n$
\For{i=1 to n}
    \If {$\AS = \emptyset$}
        \State Set all remaining $y_k = 0$
        \State \Return $y$
    \EndIf

    \If {$\exists y_1, x_1, x_2$ s.t. $y_1 = f(x_1, x_2)$}
        \State $\AS_0, \AS_1 = \textsc{AffineReduce}(\AS, f)$
        \State Find $b \in \{0,1\}$ that minimizes $|\AS_b|$, Set $y_1 = b$, $\AS = \AS_b$
    \Else 
        \State Find  $y_1, y_2, x_1, x_2, x_3, x_4$ s.t. $y_1 = f_1(x_1, x_2, x_3)$, $y_2 = f_2(x_2,x_3, x_4)$
        
        \State Find $b_1, b_2 \in \{0,1\}$, s.t. 
         \\ \hspace{1.8cm} $A = \{ (x_1, x_2, x_3, x_4) \in \F_2^4 \colon (f_1(x_1, x_2, x_3), f_2(x_2, x_3, x_4)) = (b_1, b_2)\}$ and $|A| \leq 4$
        \State Let $\AH\subseteq\F_2^4$ be the hyperplane defined by points in $A$ 
        \State Let $\AH' = \{x \in \F_2^n \colon (x_1,x_2,x_3,x_4) \in \AH\}$
        \If{$\AS \nsubseteq \AH$}
            \State Set $(y_1, y_2) = (b_1,b_2)$, $\AS = \AS \cap \AH'$

        \Else
            \If{$\exists(c_1, c_2)$ s.t. $\forall (x_1, x_2, x_3, x_4) \in \AH$ $(f_1(x_1, x_2, x_3), f_2(x_2, x_3, x_4)) \neq (c_1, c_2)$}
                \State Set $\AS = \emptyset$, $(y_1, y_2) = (c_1, c_2)$

            \Else
                \State Find $(c_1, c_2)$ s.t. \\ \hspace{3cm} $\AU = \{(x_1, x_2, x_3, x_4) \in \AH \colon (f_1(x_1, x_2, x_3), f_2(x_1, x_2, x_3)) = (c_1,c_2) \}$,$ |\AU| \leq 2$
                \State Let $\AU' = \{x \in \F_2^n | (x_1,x_2,x_3,x_4) \in \AU\}$

                \If{ $\AS \nsubseteq \AU'$
                }
                    \State Set $\AS = S \cap \AU'$, $(y_1, y_2) = (c_1, c_2)$

                \Else
                    \State $\AS = \emptyset$, $(y_1, y_2) = (1-c_1, c_2)$
                    
                \EndIf

            \EndIf
        
        \EndIf
    \EndIf
    
\EndFor

\State \Return $y$
\end{algorithmic}
\end{algorithm}

\end{document}